# Annotations for HTML to VoiceXML Transcoding: Producing Voice WebPages with Usability in Mind

**Zhiyan Shao**
**Robert Capra**
**Manuel A. Pérez-Quiñones**
Department of Computer Science
Virginia Tech
Blacksburg, VA
Emails: {zshao | rcapra | perez}@vt.edu

## ABSTRACT

Web pages contain a large variety of information, but are largely designed for use by graphical web browsers. Mobile access to web-based information often requires presenting HTML web pages using channels that are limited in their graphical capabilities such as small-screens or audio-only interfaces. Content transcoding and annotations have been explored as methods for intelligently presenting HTML documents. Much of this work has focused on transcoding for small-screen devices such as are found on PDAs and cell phones. Here, we focus on the use of annotations and transcoding for presenting HTML content through a voice user interface instantiated in VoiceXML. This transcoded voice interface is designed with an assumption that it will not be used for extended web browsing by voice, but rather to quickly gain directed access to information on web pages. We have found repeated structures that are common in the presentation of data on web pages that are well suited for voice presentation and navigation. In this paper, we describe these structures and their use in an annotation system we have implemented that produces a VoiceXML interface to information originally embedded in HTML documents. We describe the transcoding process used to translate HTML into VoiceXML, including transcoding features we have designed to lead to highly usable VoiceXML code.

## INTRODUCTION

Telephone-accessible voice interfaces to information stored on the web are becoming more prevalent. Services such as TellMe and BeVocal provide voice access to news, sports, weather, financial and entertainment information using VoiceXML-based interfaces that are accessible from any telephone. However, much of the web-based information that is currently available through voice interfaces is limited in scope and has been selected and processed by the voice service providers to integrate into existing voice information systems that have been crafted for usability. A wide variety of existing web content exists in the form of HTML web pages.

Transcoding can be used to convert an HTML document (e.g. a web page) to a VoiceXML document (e.g. a voice interface). Transcoding is a method for translating one type of code (e.g. HTML) into a different type (e.g. VoiceXML). Transcoders exist to convert HTML to VoiceXML [12], to convert HTML to a format more suitable for display on a PDA [8][2][5], and to convert from a UIML document to multiple language documents [1][14].

The process of converting HTML code into VoiceXML code is a fairly straightforward application of *transcoding*. However, a difficulty arises not in *how* to translate from the original HTML to the target VoiceXML – but instead in *what* the resulting VoiceXML code should be? For our work, we focus on transcoding an HTML page to produce a VoiceXML page with high usability.

Transcoding the diverse set of "raw" information stored in HTML web pages into VoiceXML is a significant challenge. First, the problem is difficult because much of the information stored on the web has been specifically engineered for use by graphical web browsers. The serial nature of voice interfaces requires different presentation strategies than the high-bandwidth, parallel nature of graphical interfaces. Furthermore, there is semantic information that is implicitly encoded in the page contents can be difficult to obtain.

The problem of converting HTML-to-VoiceXML has some similarities to the problem of displaying HTML on small screen devices. The small screen-size constrains the amount of data that can be presented "at once" in a similar way that the serial nature of voice limits data presentation. Research on web page presentation for small screen devices has explored ways to summarize information and to divide web pages into logical and physical-layout "segments" that can be used to control the presentation of large or complex pages. However, the small screen devices still rely on the visual scanning capabilities of humans as desktop web browsers do, even if the screen size is smaller. The HTML-to-VoiceXML problem has an extra restriction -- the serial (and thus slow) nature of voice presentation.

In this paper we present the results of our research into building a servlet to implement part of a transcoder that produces VoiceXML pages from HTML files with external annotations. The servlet relies on semantic tags that are added to an HTML file to drive the generation of a VoiceXML interface with high usability. For our initial work, we focused on structured web pages, such as those available in news sites, like CNN.com or My Yahoo! Our assumptions for this work are not that users will abandon their graphical web browsers in favor of voice web browsing, but rather that mobile users will be able to gain directed access to information stored on HTML web pages through a usable, transcoded voice interface.

## RELATED WORK

The IBM WebSphere Transcoding Publisher (WTP) [12] is a commercial product that supports an HTML-to-VoiceXML transcoder. Transcoders can be plugged into a WTP server that can be configured as a proxy. Client-side browsers can use this proxy to obtain transcoded content. The proxy intercepts the HTTP from the client-side browser, fetches the requested HTML document, transcodes it, and forwards on the transcoded document to the browser. In the case of the HTML-to-VoiceXML transcoder, a voice browser could be configured to use the proxy to receive VoiceXML versions of HTML web pages. IBM has an HTML-to-VoiceXML transcoder for use with WTP that splits the HTML into two main sections in the VoiceXML code it produces: a main content section and a listing of all the links on the page [7]. In addition, a menu is added to the VoiceXML file to allow users to navigate between the two other sections, or to exit.

The main content section contains the text from the web page. The transcoder divides the main content into subsections based on heading tags (e.g. <h1>, <h2>, etc.). The text between heading tags is used to create menus and speech recognition grammar choices in the VoiceXML code. Users can speak any of the prompted choices to listen to the paragraphs following the heading tags. The link list section contains a listing of all the links available on the page. The text between the HTML link tags (e.g. <a></a>) is used in the speech recognition grammar choices and as the text to be read in the list of link choices.

This transcoder has a simple translation strategy that works for simple HTML documents with clearly structured heading tags and for documents that the text between the <a></a> tags is context independent. Unfortunately, most web pages do not have these characteristics. For example, if the transcoder translates a page that has no heading tags in it, the result is of very low usability. The resulting VoiceXML code presents all the text on the page sequentially without any user control.

Another approach that solves some of the problems of automatic transcoding is one that uses annotations. Two papers [2][8] describe annotation-based transcoding research done at the IBM Tokyo Research Laboratory. In both of these cases they use the Transcoding Architecture within the WTP. Hori et al. [8] designed a system to make HTML documents suitable for small-screen devices. Asakawa et al. [2] used the same external annotations method and organize visually separate sections of an HTML page so they can be presented together in a voice presentation. Both projects focused on using annotations to mark the importance value of an HTML page subsections and to then reconstruct the page for browsing in a PDA or a voice browser. Our approach relies on external annotations for the first phase (Phase I) of transcoding. In Hori [8] and Asakawa [2] annotations are used to highlight sections of pages that are of interest for later processing. We use annotations to remove unneeded data and to insert our own new tags. The result is an XML file with information for a voice interface. Then in Phase II of the transcoding, we use a servlet transcoder to translate this XML stream into VoiceXML output.

The Aurora transcoding system [9] has some similarities with our work. They adapt web pages based on semantic information and build an XML document with extracted information. Their intent is for the transcoded document to support navigation in Internet Explorer and in IBM Homepage Reader. Their work is more geared towards specific tasks that users may perform, such as interaction with an auction site. The semantic information used for the transcoding must be produced manually.

Speech Application Language Tags (SALT) [16] is a project that is in its early stages. SALT tags are added to an HTML document so that users with a special browser can interact with the Web using graphics and voice at the same time. SALT is different from VoiceXML in several ways. First, it is intended to extend the existing web browser with a voice interface, and not as an alternative interface style. Also, SALT takes a programming approach to adding voice to the web, while VoiceXML uses a document-based approach. SALT applications are composed of objects, triggers and events, while VoiceXML applications are built by combining tags into one or more documents.

XHTML+Voice [19] has an approach that is similar to SALT, with a focus on supporting multi-modal devices. In this approach, existing VoiceXML tags are integrated into XHTML. In contrast, SALT tries to integrate new tags into HTML.

Several Interfaces, Single Logic (Sisl) [3] is an architecture and domain-specific language (DSL) for single service logic to support multiple user interfaces. It provides a high-level abstraction of the user-system interaction. The idea is to design the transaction to be provided in DSL and then to convert it into several interfaces (including, for example, VoiceXML and HTML). Sisl is useful when trying to build a system from scratch that will support multiple interfaces. However, it does not support converting existing HTML pages to VoiceXML.

UIML is a language designed to build multi-platform interfaces. The language relies on a number of transformations, similar to transcoding, that changes code from a high level representation (platform independent) to a platform specific representation. Farooq et al. [1] have developed transformations to convert generic UIML to platform-specific UIML. This work can also be included in the same category with Sisl, in that they both can be used to create multi-platform interfaces from scratch, but cannot help in the transcoding of existing content. Others have used UIML to produce voice interfaces [14].

Emacspeak [15] uses information from an HTML file to support auditory interfaces. It obtains its data by analyzing HTML tags, such as h1, h2. The resulting file is motivated by the needs of visually impaired users. With the new auditory version, users can listen to the prompt and respond making selections through keyboard.

Tell-me and BeVocal [17][4] are two centralized services that provide access to information that is often found on the web via a phone voice user interface. The personnel at either of these organizations organize the information gathered from the web in a form that a suitable for use in a voice user interface. This approach has higher usability than the transcoding approach but the information available is restricted to only those web pages that the central service has "converted." They take advantage of VoiceXML to provide a more flexible and economic voice service than with traditional interactive voice response (IVR) systems.

Some transcoding research has also been done for PDAs to provide access to the WWW based information on handheld devices. Orkut et al. [5] divide web pages using the syntactic structure of HMTL pages. Using HTML tags like <p>, <ol> and <table>, they split a Web page into several units and create multiple representations of it (incremental, keyword and summary), to compactly display these units on a PDA. Their worked allows the user to have an overview of a web page stored in their PDA and to explore interesting parts of the page. Some strategies, like displaying first few words of a sentence and progressively expanding to more content, fit well on small screens using pen-based input interaction. However, many of their ideas are not applicable for voice interaction.

## OUR APPROACH

Automatically transcoding an HTML file to produce a VoiceXML file is a difficult problem, particularly if we are concerned about the usability of the resulting voice user interface (VUI). HTML, and the interfaces it describe, has been designed with a structure optimized for visual scanning. In many cases, web pages contain sections that might not need to be transcoded when presenting the page by voice. For example, menus often appear both at the top and bottom of web page. Clearly these do not need to be presented twice to the user in a voice user interface. Images and banners that show most of their content graphically can be difficult, if not impossible, to communicate via voice. Extra information that is often presented in sidebars may be appropriate for inclusion in a graphical display than for inclusion when presenting a page using a serial technique such as voice output.

Links also present a problem for transcoding HTML to voice interfaces. One of the advantages of hypertext is the ease with which you can include in the document many connections/links to other information. Our visual perceptual system sees these without much effort, and the user decides if he/she needs to follow any of these links. However, for a VUI, presentation of links and selection of items is less straightforward. First, it is difficult to present the link to the user at the same time that the textual content of the link is being presented. In graphical web browsers, links often appear underlined or with some other graphical form of feedback, but in a VUI links need to be audibly identified. Some approaches used include: changing the voice of the presentation for the link (e.g. male to female) [18], changing the pitch or volume, or even producing some background sound to indicate that the words being read are part of a link [11]. Which of these to use is still a research problem, and the location of the link in the sentence seems to have an effect on its effectiveness [18][6].

Furthermore, contextual information is often important to users' understanding of hyperlinks. In some cases, links can be almost meaningless without the context where they occur. For example, many web pages include links to organizations when they are referred to in a story (e.g. when IBM is mentioned in a story, a link to www.ibm.com is used). These links often do not provide additional details specific to the story, but simply lead to the main page of the company or a related web site. For the visual reader, he/she can expose the URL by moving the mouse pointer over the link and quickly decide if the link is one they wish to follow. This operation on the graphical user interface is simple, relies on a simple UI action (move the mouse), and uses the user's "general knowledge" of how the web works. Presenting this type of contextual information for links in a voice interface is more challenging.

Web pages may also contain links to other types of related information such as relevant news stories or links for navigation (e.g. click here for more). Often these types of links require semantic knowledge to determine if they should be presented in a voice interface or if they should be suppressed to help streamline the presentation of the main content.

Another source of problems with links is the text to be presented to the user and the expected response from the user to activate the link. The text in an <a href> tag on the web page may not be well suited for use in a voice interface. There are several problems here. First, the link text on the web page may be underspecified or ambiguous. Consider web pages that include items such as "For more information click here" – where the word "here" is the hyperlink text contained in the <a href> tag. In this example, the word "here" alone does not provide much contextual information about the link and could easily become ambiguous if there was another "here" link on the same page. One HTML to VoiceXML transcoder we reviewed used this approach to create a menu of links available on the page based (and presented) solely on the link text. Because of the issues mentioned above, the resulting menu of link choices was difficult to understand and could result in ambiguous choices.

At the other extreme, HTML link text can, be overspecified for use in a voice interface. If the link contains a long sentence, such as is typical in news sites (e.g. "Company Executives Deny Pressure to Meet Targets"), it is not clear what users will say to select this option. Should we define all sub-phrases of the link text as grammar choices (e.g. Company, Company Executives, Company Executives Deny, etc.)? Or should we have single word choices for all words in the title? Requiring that the whole link to be spoken seems unreasonable and it places a heavier burden on short-term memory; we found this to be a source of errors and frustration in our usability evaluation [13]. Another options is to use a link independent method to select the choice (e.g. the user saying "this one"), but this approach may have issues related to the timing of the prompt and the users speech.

Given the slow presentation medium that a VUI relies on, we need to optimize the interface and thus avoid presenting more that the information required. For the reasons presented above, we reached the conclusion that we need to be selective on what links to include in a transcoded VUI.

With these ideas in mind, we set out to determine what would be the appropriate VUI for a structured web page, much like the My Yahoo! page. We constructed an initial VUI and performed some iterative refinement to validate our design. Details of this evaluation can be found elsewhere [13]. The following sections describe our current thinking about what this VUI ought to be like based on structures we have observed that occur frequently on web pages and present a resulting technology implementation to produce the desired VUI.

## STRUCTURES OF WEB PAGES

We (and others) have observed that web pages tend to have several types of regular structures that can be identified and that are useful in presenting the data contained on the page.

The concept of design patterns can be applied to web page structure to identify repeated patterns and structures that appear on web pages.

As an informal starting point for our research, we located, downloaded, and printed a set of about 50 web pages and investigated their structure. From this set, we identified an initial set of four graphical web page structures that have different characteristics for translation to voice presentation. Each of these structures is described in the sections below.

### Paragraphs and Free-form text

This is general text that is presented like text in a book, newspaper, or article. This text may contain detailed information and is intended to be read carefully. Voice presentation can be accomplished by using a text-to-speech synthesizer to read the text sequentially.

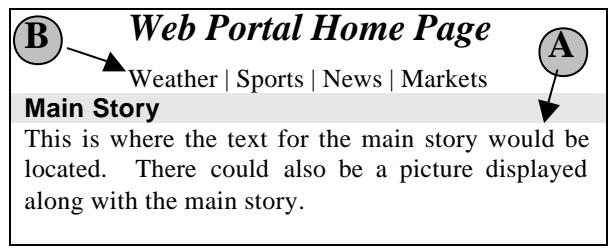

**Figure 1: Example Web Portal Page (top)**

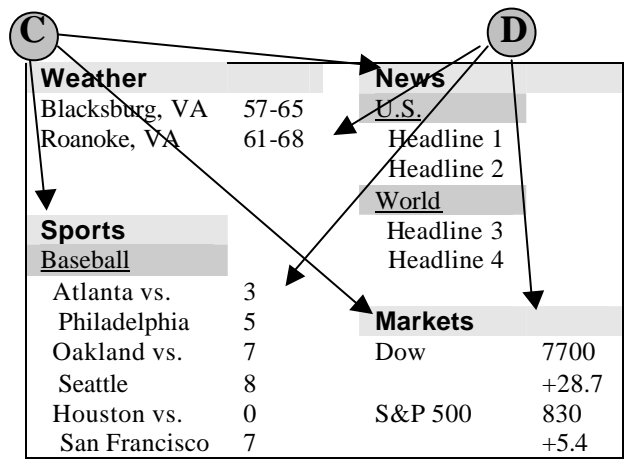

**Figure 2: Example web portal page (bottom)**

The type of controls to be used in the VUI presentation of free-form text are similar to those available in screen readers. It will be important to give users control of the speed of presentation and usable controls for navigating forward and backward through the text. In Figure 1, the element marked by a circled "A" is an example of paragraphs/free-form text.

Free-form text (and other structures) could contain HTML code with hyperlinks that would need to be presented to the user. Several techniques for presenting hyperlinks to users are compared and evaluated in [18].

### Hierarchical Menu Structure

Many web pages present information in structured components and sometimes these components have a hierarchical menu structure. There are several ways such a menu structure can occur in an HTML document. First, the structure may be explicitly indicated through HTML elements such as headings (e.g. <h1>, <h2>, etc.), frames, or tables. In other cases, the menu structures may be more implicitly defined by sequences of elements such a list of links embedded in the document (for example, the elements marked with the circled "B" in Figure 1 form an implicit menu structure). These explicitly and implicitly created hierarchical structures on web pages can be used to help improve a user's ability to quickly navigate to sections of a page in a voice interface.

**Repeated, Structured Information**

This type of information is one that has a clear structure (such as fields in a data-base record) and that is repeated several times in a web page. Figure 2 gives several examples of repeated structured information marked with the letter "D". For example, the weather section includes a repeated structure that has the city name, state abbreviation, and the low and high temperature. This type of information is likely to contain one or more key fields that may be used to access the other information (such as the city name or market index name).

There are several features that can be used to help voice navigate lists of repeated, structured information. Each "record" in the list could be read completely (i.e. all its fields) in sequential order. Navigation commands could be provided to allow users to move forward and backward in the list, including the ability to "barge-in" to issue a command. Additionally, users could be allowed to request only one specific item. For example, a user may wish to navigate to a specific movie in a list by saying its name. While the specific movie information is being read, the user could then say a field name such as "starting times" to navigate directly to that part of the record. There are many user interface issues to be explored regarding features for presenting this type of information.

**Regions**

Many web pages have a structure that divides the page into visual regions of related content. In many cases, each of these rectangular regions is independent (almost) of the other content on the page and could be considered a "sub-page" of the original page. Each region is like its own mini-web page that can, in turn, have its own structures (i.e. free text, menus, repeated structures, more frames).

Web page designers design these regions to be visually distinct when displayed in a graphical web browser. In our voice interface, we want to attempt to preserve (and translate) this design of the visual web page into the voice presentation of the page. This can be accomplished by presenting these regions as sub-menus of the main page, thus allowing quick access to that area of the original web page by selecting the heading for that section from a menu of choices. This effectively translates the single original web page into a hierarchical set of sub-pages linked through menus in much the same way that the hierarchical menu structure described above works

Supporting navigation through these sub-pages requires adding the ability for users to go "back" up the hierarchy or "back" through their navigation history. Additionally, there may be links that are prominent on the original page that are moved deeper into the hierarchy of sub-pages upon translation. This could cause confusion for users who are accustomed to using a graphical version of the web page – they may think of the page in terms of directly navigating to information by looking at a particular heading or clicking on a particular link that gets buried several layers in the translated voice interface. To help improve the usability of such pages, we have incorporated the ability to "promote" items from lower in the sub-page hierarchy to higher levels through the inclusion of those items in the voice grammar of the higher-level sub-pages.

**Voice Navigation Model Desired**

It is not clear to us that there is an "equivalent" VUI design to a given HTML page design. Nevertheless, we need to be able to generate a VUI given an HTML page. We opted to use external annotations to extend the HTML with semantic information that would lead to a VUI with good usability. We designed the tags used in the annotation file based on: 1) our observations of the structure of web pages, 2) principles of prompt design for VUI [20], and 3) our usability evaluations of early prototypes [13].

VUIs, particularly those built for use over a phone, often rely on a hierarchical menu structure where the user is moving down a menu tree until he/she reaches a leaf node. It is at the leaf node that information is presented. This structure is very effective for novice users; it amounts to a system directed dialogue of the options available in the system. The VUI does not need to present all menus available at each level, although it often accepts all choices. We can use incremental prompts and expanded prompts [20] to speed up the interaction while still providing support for novice users.

As we discussed in the previous section, there is often a clear hierarchical structure in web pages that can be used to help navigate within a web page using a VUI. We make use of this structure to build the menu tree for the VUI. This hierarchical menu structure allows us to define a "back" command that presents to the user a similar navigation model to the model used by the graphical browsers.

To support advanced users, we allow for lower levels of the menu tree to be "moved" up to higher levels. This is done to support experienced users by giving them direct access to information without the requirement to navigate the full tree. We call this process "promotion", because lower level menu items are promoted to allow access from higher-level menus. The "Transcoder Design" section below explains issues involved in performing this transcoding.

Finally, it is worth restating our assumptions of how users will use our VUI. We do not believe that users will stop using their Web browser in favor of a telephone-based voice user interface. We think that users will use the telephone voice interface to have directed access to particular pieces of information. This is similar to the approach done by BeVocal and TellMe, where they provide quick and direct access to information of general interest, such as news, weather, sports, movies, stocks, etc.

**IMPLEMENTATION ARCHITECTURE**

Our implementation is built using IBM's WebSphere Transcoding Publisher (WTP) system. Details about the IBM WTP system can be found in [10].

Figure 4 shows an overview of our implementation. Transcoding using the IBM WTP can be viewed in two main phases. The first phase (marked with an "I." in Figure 4) involves using an annotation transcoder to insert tags and clip regions of the original HTML document, leaving only the parts that we want to convert to VoiceXML. This phase can utilize an external annotation file that indicates what should be inserted and removed from the original HTML document. Positions of elements to be inserted in the document are specified in terms of their XPATH. The result of this phase is a document that uses an XML-based language we have developed called VXPL (VoiceXml Precursor Language) that will be described in more detail in the next section.

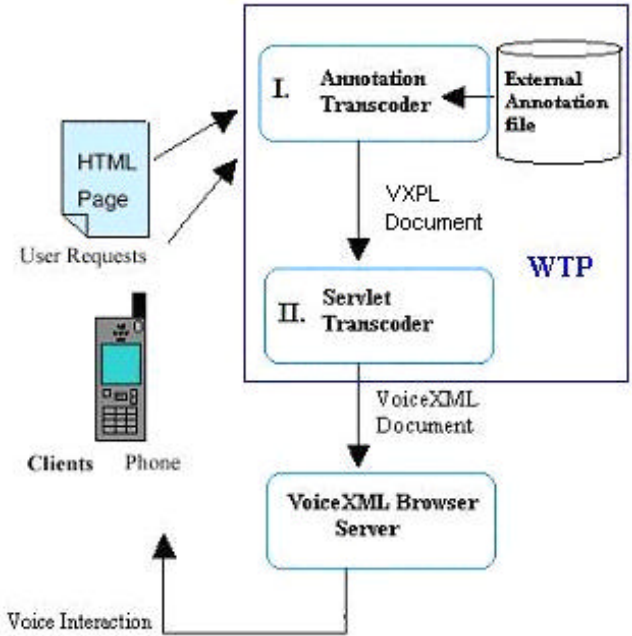

**Figure 4: Transcoding Architecture**

Before the annotations are processed, the HTML document is parsed into a document object model (DOM), where it is represented as a tree where each node in the tree is a tag in the original HTML. The XPATH of a node is a unique path to that node in the DOM tree. XPATHs are used in the external annotation file to specify where elements should be inserted from the original HTML document [10]. This type of reference method is especially useful for structured web pages, such as web portals and news sites. Many of these types of sites have a relatively fixed structure of tables, headlines, and sidebars. The content of these pages may change frequently, but in many cases these pages are generated by filling in a template document with specific content. Since the templates are usually used for an extended period of time, the basic structure changes much less frequently than the content.

The second transcoding phase (marked with an "II." in Figure 4) involves using a Java servlet to translate the VXPL document into VoiceXML. Our current work mainly focuses on this second phase, transcoding the VXPL document into VoiceXML. The WTP architecture allows new transcoders to be added to the system as Java servlets. Our servlet transcodes the VXPL into VoiceXML and adds additional functionality to support usability features such as navigating "back". Usability features added to the VoiceXML code will be described in more detail later in this paper.

## VXPL Tags

We developed an XML language called VXPL based on the structures we observed in our examination of web pages and the desired navigation model. In this section, we describe how VXPL relates to our observations and how each VXPL tag can be used.

The goal of VXPL is to facilitate the translation of HTML to VoiceXML. Content is extracted from the HTML document and embedded into a VXPL counterpart. We have designed VXPL to allow both manual and automatic creation of VXPL code from HTML.

VXPL provides tags that can be used to indicate structures in the HTML page that are useful for creating a voice interface to that page in VoiceXML. For example, the <menu> and <menuitems> tags are used to indicate hierarchical menu strucutures. The <prompt> tag is used for notating text that should be played as a prompt. The tags <structured> and <field> provide a way to describe repeated structured information such as the weather example from Figure 2.

```
<root>
<menu id="top">
    <prompt>Welcome to the Web Portal</prompt></menu>
<menu id="news" parentid="top"> <prompt>News </prompt></menu>
<menu id="us" parentid="news"> <prompt>US </prompt></menu>
<menuitem parentid="us">
    <a href="url for Headline 1">Headline 1</a>
</menuitem>
<menuitem parentid="us">
    <a href="url for Headline 2">Headline 2</a>
</menuitem>
<menu id="world"><promp>World</prompt></menu>
.....
<menu id="weather" parentid="top">
    <prompt>Weather</prompt></menu>
<structured id="bburg" parentid="weather" />
    <field key="true" parentid="bburg">Blacksburg, VA</field>
    <field name="Low" parentid="bburg">57</field>
    <field name="High" parentid="bburg">65</field>
<structured id="roanoke" parentid="weather" />
.....
<menu id="sports" parentid="top">
    <prompt>Sports</prompt></menu>
<menu id="baseball" parentid="sports" promote="true">
    <prompt>Baseball</prompt></menu>
<structured id="Atlanta-Philly" parentid="baseball" />
    <field key="true" parentid="Atlanta-Philly">Atlanta</field>
    <field key="true" parentid="Atlanta-Philly">Philadelphia</field>
    <field name="Atlanta" parentid="AtlantaPhilly">3</field>
    <field name="Philadelphia" parentid="AtlantaPhilly">5</field>
<structured id="Oakland-Seattle" parentid="baseball" />
.....
</root>
```

**Figure 5: VXPL Sample Code**

Figure 5 represents an example of our current implementation of VXPL. This example is based on the web page section shown in Figure 2. We are continuing to refine and evolve the VXPL specification as we continue our research.

## TRANSCODER DESIGN

Our transcoder focuses on translating VXPL into VoiceXML code. Figure 8 shows a partial example of transcoding the VXPL in Figure 5 into VoiceXML using our transcoder. Figure 10 illustrates the resulting menu structure for the VoiceXML code that is created by the transcoder.

Since VXPL is an XML document, the Phase II transcoding (from VXPL to VoiceXML) is basically a tree translation. We chose to use a Java-based XML parser that allows manipulation of the DOM to accomplish the translation because it seemed well suited to our needs. However, other XML translation techniques such as the use of XSLT might also work as a way to translate the VXPL to VoiceXML.

```
<var name="aNavHistory" expr="new Array()" />
<var name="sDefaultURL" expr="#top" />
<catch event="ongoback">
  <script>aNavHistory.pop()</script>
  <var name="sJumpTo" expr="(aNavHistory.pop())
       ||application.sDefaultURL " />
  <goto expr="sJumpTo" />
</catch>
<form id="top">
  <block> <script>aNavHistory.push("#top")</script>
       <goto next="#topmenu" /> </block>
</form>
<menu id="topmenu">
  <prompt>
       Welcome to the Web Portal. Please say one of the
              followings: <enumerate/>
  </prompt>
       <choice next="#news">News</choice>
       <choice next="#weather">Weather</choice>
       <choice next="#sports">Sports</choice>
       <choice next="#baseball">Baseball</choice>
</menu>
<menu id="newsmenu">
......
  <choice event="ongoback">back</choice>
</menu>
```

**Figure 8: Partial Transcoded VoiceXML**

Two main features are added during the transformation. First, lower-level menu items can be promoted to the highest-level menu (main menu) if the *promote* attribute of the *<menu>* tag has been set to true. This is illustrated by the "baseball" VXPL menu in Figure 5. This menu item has the *promote* attribute set to *true*, so when it is transcoded the resulting VoiceXML main menu will contain a *<choice>* item for "baseball" (see Figure 8).

Second, a "back" choice is added to each menu in the VoiceXML code (except for the main menu). This is accomplished by adding a grammar item "back" to each menu that, if spoken, will result in a "back" event. JavaScript code (shown at the top of Figure 8) is added to each page to manage a history stack that controls what page to go back to when a back event is raised.

The combination of a "back" choice and the promote feature created an issue about where to go back to when a user navigated using the promoted choice. For example, in Figure 10, what should happen if the user gets to Baseball from the main menu (Level 1) and then says "back"? It doesn't make sense to go back to the Sports (Level 2). Instead, the system should go back to main menu.

We use the JavaScript history stack to record the user's selection for every menu selection. When user says "back", the VoiceXML document throws an "ongoback" event. In the event handling, this pops the top menu choice off the stack and jumps to that menu using a VoiceXML <goto>. There is a push function that is added before each menu to push this menu's id into the stack. Through the use of the "push" and "pop" operations, we accomplish the implementation of a "back" choice.

| Level 1 | Level 2 | Level 3 | Level 4 |
|---|---|---|---|
| Welcome to Web Portal | | Blacksburg, VA | |
| | | Roanoke, VA | |
| | Sports | Baseball | Atlanta |
| | | | Oakland |
| | | | Houston |
| | News | U.S. | Headline 1 |
| | | | Headline 2 |
| | | World | Headline 3 |
| | | | Headline 4 |
| | Markets | Dow | |
| | | S&P 500 | |
| | Baseball | Atlanta | |
| | | Oakland | |
| | | Houston | |

**Figure 10: Menu Structure of VoiceXML document**

## DISCUSSION AND FUTURE WORK

The use of cell phones is extending the reach of web-based information and enhancing the effectiveness of mobile information workers. Unfortunately, there is no easy way to transcode the content available on the web into a voice user interface that has good usability. Web pages have been designed for graphical displays and this makes automatic transcoding to a voice user interface a very difficult problem.

In our research, we explored the types of structures available on web pages. For this set, we defined the desired voice interaction style for each structure. We then defined a set of XML-based tags that are used in a transcoding architecture to generate a VoiceXML document that produces the desired behavior. The ultimate goals is that we can define external annotation files for existing HTML files to make these pages available over a phone-based voice user interface.

In the process of studying this domain, we ran into some interesting problems that we have begun to address further. The first is how to present the menu of news stories to the user. News stories are often long headlines, and expecting the user to repeat the whole store is unrealistic and error prone. But shortening the news story headline might introduce ambiguities in the resulting grammar as there might be more than one choice with the same set of words. Besides, the user will not know what is the appropriate subset of the headline to say.

For now, we have added a number in front of the story, so all stories have a unique number, and the user selects the story by saying the number. We will be exploring other options in a usability study soon.

Another limitation of our current system is in phase 1. The current external annotation transcoder had a bug that prevented us from completely implementing phase 1 of the system. The transcoder we used can only insert a start tag and end tag at the same position (e.g. <menu></menu>). We needed to insert the start tag in one location of the file and the end tag after some of the text in the HTML file. We are exploring different annotation engines for our next round of implementation.

Finally, the text tag has not been implemented yet. This tag, as a minimum, simply plays the text included within its start and end tag. This implementation is straight forward. More interesting will be to include presentation controls inside of the <text> so the user can speed up or slow down the textual presentation by using barge-in commands. We need to conduct some usability evaluations with the VUI style that we design before we implement it, to ensure a good usability in the resulting interface.

## ACKNOWLEDGMENTS

The authors wish to thank Natasha Dannenberg for her initial research on the usability of VUI that lead to many of the ideas presented here. We also wish to thank Dr. Naren Ramakrishnan for the many informal but insightful conversations regarding the ideas here presented. This research was funded in part by an IBM Grant to explore the use of VoiceXML within the WebSphere Transcoder line of products.

## REFERENCES

1. Ali, M.F., Pérez-Quiñones, M.A., Abrams, M. and Shell, E. Building Multi-Platform User Interfaces with UIML, in Proceedings of CADUI 2002, May 2002.

2. Asakawa, C. and Takagi, H. Annotation-based transcoding for nonvisual web access. The fourth international ACM conference on Assistive technologies ASSETS. 172-179. ACM. 2000.

3. Ball, T., Colby, C., Danielsen, P., Jagadeesan, L.J., Jagadeesan, R., Laufer, K., Mataga, P. and Rehor, H. Sisl: Several Interfaces, Single Logic, International Journal of Speech Technology 3, 93-108, 2000.

4. Bevocal. Available at http://www.bevocal.com

5. Buyukkokten, O., Kaljuvee, O., Garcia-Molina, H., Paepcke, A. and Winograd, T. Efficient web browsing on handheld devices using page and form summarization, ACM Transactions on Information Systems (TOIS) January 2002, V20, N1.

6. Christian, K., Kules, B., Shneiderman, B. and Youssef, A. A Comparison of Voice Controlled and Mouse Controlled Web Browsing. ASSETS 200, November 13-15, 2000. Washington.

7. Hopson, Nichelle. WebSphere Transcoding Publisher: HTML-to-VoiceXML Transcoder. January 2002. Accessed on October 4, 2002 from: http://www7b.boulder.ibm.com/wsdd/library/techarticles/0201_hopson/0201_hopson.html

8. Hori, M., Kondoh, G., Ono, K., Hirose S.I. and Singhal S. Annotation-Based Web Content Transcoding, Proceedings of the 9th International World Wide Web Conference. Amsterdam, Netherlands: 1999.

9. Huang, A.W. and Sundaresan, N. Aurora: A Conceptual Model for Web-Content Adaptation to Support the Universal Usability of Web-based Services, CUU '00 Arlington VA.

10. IBM Corporation. WebSphere Transcoding Publisher Version 4.0 Developer's Guide. Available at: http://www-1.ibm.com/support/docview.wss?uid= swg27000533&aid=1

11. James, F. Lessons from Developing Audio HTML Interfaces, ACM Conference on Assistive Technologies. April 15-17, 1998, Marina del Rey, CA USA, pages 27-34.

12. Lamb, M. and Horowitz, B. Guidelines for a VoiceXML Solution Using WebSphere Transcoding Publisher. Available at http://www-3.ibm.com/software/webservers/transcoding/library.html

13. Pérez-Quiñones, M.A., Dannenberg, N. and Capra, R. "Voice Navigation of Structured Web Spaces", Technical Report TR-02-22, Computer Science Department, Virginia Tech.

14. Plomp, C.J. and Mayora-Ibarra, O. A. Generic Widget Vocabulary for the Generation of Graphical and Speech-Driven User Interfaces, International Journal of Speech Technology 5, 39-47, 2002.

15. Raman, T.V. Emacspeak- Toward The Speech-enabled Semantic WWW. Available at http://www.cs.cornell.edu/Info/People/raman/publications/semantic-www.html.

16. SALT. Available at http://www.saltforum.org/.

17. Tellme. Available at http://www.tellme.com.

18. Wang, Q.Y., Shen, M.W., Shi, R.D. and Su, H. Detectability and Comprehensibility Study on Audio Hyperlinking Methods. Proceedings of the Human-Computer Interaction - Interact '01, pp. 310-317.

19. XHTML+voice. Available at http://www.w3.org/TR/xhtml+voice/.

20. Yankelovich, N. How do Users Know What To Say? ACM Interactions, Volume 3, Number 6, November/December 1996.